\documentclass[conference]{IEEEtran}
\usepackage{color, colortbl}
\usepackage{tabu}
\usepackage{xcolor}
\definecolor{Gray}{gray}{0.6}

\usepackage{cite}

\usepackage{amsmath}
\usepackage{graphicx}
\newif\ifshowcomments
\showcommentsfalse
\ifshowcomments
\newcommand\zxie[1]{\textcolor{red}{[zxie: #1]}}
\newcommand\fixed[1]{\textcolor{blue}{[zxie: #1]}}
\else
\newcommand\zxie[1]{}
\newcommand\fixed[1]{}
\fi

\IEEEoverridecommandlockouts\IEEEpubid{\makebox[\columnwidth]{978-1-5090-1629-7/16/\$31.00~\copyright~2017 IEEE \hfill} \hspace{\columnsep}\makebox[\columnwidth]{ }}

\begin{document}
\title{Improving Palliative Care with Deep Learning}

\author{\IEEEauthorblockN{Anand Avati\IEEEauthorrefmark{1},
Kenneth Jung\IEEEauthorrefmark{2},
Stephanie Harman\IEEEauthorrefmark{3}, 
Lance Downing\IEEEauthorrefmark{2},
Andrew Ng\IEEEauthorrefmark{1} and
Nigam H. Shah\IEEEauthorrefmark{2}}
\IEEEauthorblockA{\IEEEauthorrefmark{1}Dept of Computer Science, Stanford University\\
Email: \{avati,ang\}@cs.stanford.edu}
\IEEEauthorblockA{\IEEEauthorrefmark{2}Center for Biomedical Informatics Research, Stanford University\\
Email: \{kjung,ldowning,nigam\}@stanford.edu}
\IEEEauthorblockA{\IEEEauthorrefmark{3}Dept of Medicine, Stanford University School of Medicine\\
Email: \{smharman\}@stanford.edu}}

\maketitle

\begin{abstract}
\zxie{Title could be more specific}
Improving the quality of end-of-life care for hospitalized\zxie{hospital} patients is a priority for healthcare organizations. Studies have shown that physicians tend to over-estimate prognoses\zxie{over-estimate prognoses of life-expectancy, ... vague}, which in combination with treatment inertia results in a mismatch between patients\zxie{patient} wishes and actual care at the end of life \zxie{The effect of overestimated prognoses vague (though it's clear in intro).}. We describe a method to address this problem using Deep Learning and Electronic Health Record (EHR) data\zxie{Would start a new sentence here}, which is currently being piloted, with Institutional Review Board approval, at an academic medical center. The EHR data of admitted patients are automatically evaluated by an algorithm, which brings patients who are likely to benefit from palliative care services to the attention of the Palliative Care team. The algorithm is a Deep Neural Network trained on the EHR  data from previous years, to predict all-cause 3-12 month \zxie{all-cause 3-12 month awkward} mortality of patients as a proxy for patients that could benefit from palliative care.   Our predictions enable the Palliative Care team to take a proactive approach in reaching out to such patients, rather than relying on referrals from treating physicians, or conduct time consuming chart reviews of all patients\fixed{Abstract missing motivation of it being too time consuming to manually determine palliative care patients}. We also present a novel interpretation technique which we use to provide explanations of the model's predictions. \zxie{Could motivate interpretation technique, e.g. "To address concerns about the interpretability of suggested patients..."}
%
%

\end{abstract}


\section{Introduction}

Studies have shown that approximately 80\% of Americans would like to spend their final days at home if possible, but only 20\% do \cite{AMERICANSDIE}. In fact, up to 60\% of deaths happen in an acute care hospital, with patients receiving aggressive care in their final days. Access to palliative care services in the United States has been on the rise over the past decade. In 2008, 53\% of all hospitals with fifty or more beds reported having palliative care teams, rising to 67\% in 2015\cite{DumanovskySpecialReport}. However, despite increasing access, data from the National Palliative Care Registry estimates that less than half of the 7-8\% of all hospital admissions that need palliative care actually receive it\cite{REPORTSITE}. Though a significant reason for this gap comes from the palliative care workforce shortage \cite{Spetz2016FewRecommendations.}, and incentives for health systems to employ them, technology can still play a crucial role by efficiently identifying patients who may benefit most from palliative care, but might otherwise be overlooked under current care models.
%
%
%
%
%
%
%
%

We focus on two aspects of this problem.  First, physicians may not refer patients likely to benefit from palliative care for multiple reasons such as overoptimism, time pressures, or treatment inertia \cite{Christakis2000ExtentStudy}.  This may lead to patients failing to have their wishes carried out at end of life\cite{STANFMORT} and overuse of aggressive care.  Second, a shortage of palliative care professionals makes proactive identification of candidate patients via manual chart review an expensive and time-consuming process. 

%
%

The criteria for deciding which patients benefit from palliative care can be hard to state explicitly. 
Our approach uses deep learning \fixed{Here and for at least first half of paper machine learning is used instead of more specific deep learning in contrast to title} to screen patients admitted to the hospital to identify those who are most likely to have palliative care needs. 
The algorithm addresses a proxy problem - to predict the mortality of a given patient within the next 12 months - and use that prediction for making recommendations for palliative care referral.
This frees the palliative care team from manual chart review of every admission and helps counter the potential biases of treating physicians by providing an objective recommendation based on the patient's EHR. Currently existing tools to identify such patients have limitations, and they are discussed in the next section.


\section{Related work}

Accurate prognostic information is valuable to patients, caregivers, and clinicians \cite{KutnerInformationIllness} \cite{Steinhauser2001PreparingProviders}. Several studies have shown that clinicians are generally over optimistic in their estimates of the prognoses of terminally ill patients\cite{Selby2011ClinicianCategories} \cite{Christakis2000ExtentStudy} \cite{ViganoTheCancer} \cite{Glare2008PredictingDisease}. It has also been shown that no subset of clinicians are better at late stage prognostication than than others  \cite{White2016AExperts} \cite{Macmillan1998AccuracyHospice} . However, clinician judgment remains the most common method of predicting survival in practice \cite{White2016AExperts}. Several solutions exist that attempt to make patient prognosis more objective and automated. Many of these solutions are models that produce a score based on the patient's clinical and biological parameters, and can be mapped to an expected survival rate.   

\subsection*{Prognostic tools in Palliative Care}
The Palliative Performance Scale  \cite{Lau2006UsePrognostication} was developed as a modification of the Karnofsky Performance Status Scale (KPS) \cite{Karnofsky1948TheCarcinoma} to the Palliative care setting, and is calculated based on observable factors such as: degree of ambulation, ability to do activities, ability to do self-care, food and fluid intake, and state of consciousness. The Palliative Prognostic Score (PPS) was constructed for the Palliative Care setting as well, focusing on terminally ill cancer patients \cite{Pirovano1999ACare}.  The PPS is calculated with multiple regression analysis based on the following variables: Clinical Prediction of Survival (CPS), Karnofsky Performance Status (KPS), anorexia, dyspnea, total white blood count (WBC) and lymphocyte percentage. The Palliative Prognostic Index (PPI),  developed around the same time as PPS, also calculates a multiple regression analysis based score using Performance Status, oral intake, edema, dyspnea at rest, and delirium. These scores are difficult to implement at scale since they involve face-to-face clinical assessment and involve prediction of survival by the clinician. Furthermore, these scores were designed to be used within the palliative care setting, where the patient is already in an advanced stage of the disease --- as opposed to identifying them earlier.
%
%
\subsection*{Prognostic tools in the Intensive Care Unit}
There also are prognosis scoring models that are commonly used in the Intensive Care Unit. The APACHE-II (Acute Physiology, Age, Chronic Health Evaluation) Score predicts hospital mortality risk for critically ill hospitalized adults in the ICU \cite{Knaus1985APACHESystem.}. This model has been more recently refined with the APACHE-III Score, which uses factors such as major medical and surgical disease categories, acute physiologic abnormalities, age, preexisting functional limitations, major comorbidities, and treatment location immediately prior to ICU admission\cite{Knaus1991TheSystem}. Another commonly used scoring system in the ICU is the Simplified Acute Physiological Score, or SAPS II \cite{LeGall1993AStudy}, which is calculated based on the patient's physiological and underlying disease variables. While these score are useful for the treatment team when the patient is already in the ICU, they have limited use in terms of identifying patients who are at risk of longer term mortality, while they are still capable of having a meaningful discussion of their goals and values, so that they can be set on an alternative path of care.

\subsection*{Prognostic tools for Early Identification}
There have been a number of studies and tools developed that aim to identify terminally ill patients early enough for an end-of-life plan and care to be meaningful.

CriSTAL (Criteria for Screening and Triaging to Appropriate aLternative care) was developed to identify elderly patients nearing end of life, and quantifies the risk of death in the hospital or soon after discharge  \cite{Cardona-Morrell2015DevelopmentCriSTAL.}. CriSTAL provides a check list using eighteen predictors with the goal of identifying \textit{the dying patient}.

CARING is a tool that was developed to identify patients who could benefit from palliative care\cite{FischerACriteria}. The goal was to use six simple criteria in order to identify patients who were at risk of death within 1 year. PREDICT \cite{Richardson2015PREDICT:Directive} is a screening tool also based on six prognostic indicators, which were refined from CARING. The model was derived from 976 patients. 

The Intermountain Mortality Risk score is an all-causes mortality prediction based on common laboratory tests \cite{Horne2009ExceptionalTests}. The model provides score for 30-day, 1-year and 5-year mortality risk. It was trained on a population of 71,921 and tested on 47,458.

\fixed{The paragraph structure here seems too explicitly formulaic}

Cowen, M et al \cite{Cowen2013MortalityActivities} proposed using a twenty-four factor based prediction rule at the time of hospital admission to identify patients with high risk of 30-day mortality, and to organize care activities using this prediction as a context. One of the their motivation was to have a rule from a single set of factors, and not be disease specific.  The model was derived from 56,003 patients.

Meffert, C et al \cite{Meffert2016IdentificationScore} proposed a scoring method based on logistic regression on six factors to identify hospitalized patients in need of palliative care. In this prospective study, they asked the treating physician at the time of discharge whether the patient had palliative care needs. The trained model was then used to identify such patients at the time of admission. The model was derived from 39,849 patients.

Ramachandran, K et al \cite{Ramchandran2013ARecord} developed a 30-day mortality prediction tool for hospitalized cancer patients. Their model used eight variables that were based on information from the first 24 hours of admission, and laboratory results and vitals. A logistic regression model was developed from these eight variables and used as a scoring function. The model was derived from 3,062 patients.

Amarasingham, R et al \cite{AmarasinghamAnData} built a tool to screen patients who were admitted with heart failure, and identify those who are at risk of 30-day readmission or death. Their regression model uses a combination of Tabak Morality Score \cite{Tabak2007UsingAdjustment}, markers of social, behavioral, and utilization activity that could be obtained electronically,  ICD-9 CM codes  specific to depression and anxiety, billing and administrative data. Though this study was not specifically focused on palliative care, the methodology of using EHR system data is relevant to our work. The model was derived from 1,372 patients.

Makar, M et al \cite{MakarShort-TermData} used only Medicare claims data on older population ($\ge$ 65 years) to predict mortality in six months. By limiting their model to use only administrative data, they hypothesized an easier deployment scenario thereby making automated prognostic models more prevalent. The model was derived separately on four cohorts (one per disease type) with 20,000 patients per cohort.

\subsection*{Prognosis in the age of Big-Data}
The proliferation of EHR systems in healthcare combined with advances in Machine Learning techniques on high dimensional data provides a unique opportunity to make contributions, especially in disease prognosis \cite{Sim2016TwoMedicine}\cite{Obermeyer2016PredictingMedicine.}. All the tools described above, and those we reviewed \cite{Rose2013MortalityLearning}  \cite{Wiemken2017PredictingApproaches.} \cite{MotwaniMachineAnalysis}  \cite{CooperAnMortality} \cite{VomlelMachine}, have at least one of the following limitations. They were either derived from small data sets (limited to specific studies or cohorts), or used too few variables (intentionally to make the model portable, or avoid overfitting), or the model was too simple to capture the complexities and subtleties of human health, or was limited  to certain sub-populations (based on disease type, age etc.) We address these limitations in our work.

\section{Methods}

We approach the problem of predicting mortality from the point of view of the palliative care team by being largely agnostic to disease type, disease stage, severity of admission (ICU vs non-ICU), age etc. We take a data driven approach and build a deep learning model that considers every patient in the EHR (with a sufficiently long history), without limiting our analysis to any specific sub-population or cohort. In order to make the problem of identifying patients with palliative care needs tractable, we use the following proxy problem statement instead:

\textit{Given a patient and a date, predict the mortality of that patient within 12 months from that date, using EHR data of that patient from the prior year.}

We treat this as a binary classification problem and build a supervised deep learning model to solve it.
Other than building a model that performs well on the above problem, we are also separately interested in the model performance on a sub-problem --- the ability to predict mortality of patients who are currently admitted. This is because it is much easier for the palliative care staff to intervene with admitted patients.

\subsection*{Data Source}

STRIDE (Stanford Translational Research Integrated Database Environment) \cite{LoweSTRIDEPlatform} is a clinical data warehouse supporting clinical and translational research at Stanford University. The snapshot of STRIDE used in our work comprises the EHR data of approximately 2 million adult and pediatric patients cared for at either the Stanford Hospital or the Lucile Packard Children's hospital between 1995 and 2014.

\subsection*{Constructing a Dataset for Supervised Learning}
Patients who have a recorded date of death are considered \textit{positive cases}; other patients are considered \textit{negative cases}. Further, we define the \textit{prediction date} of a patient to be the point in time that divides their health record timeline into virtual future and past events.  We use data from each patient's virtual past to make predictions about their survival 3-12 months in the future.  Note that we must take care when defining the \textit{prediction date} to not violate common sense constraints (described below) that could invalidate the labels.  We only include patients for whom it is possible to find a \textit{prediction date} that satisfies these constraints.  


\begin{figure}
    \includegraphics[width=\columnwidth]{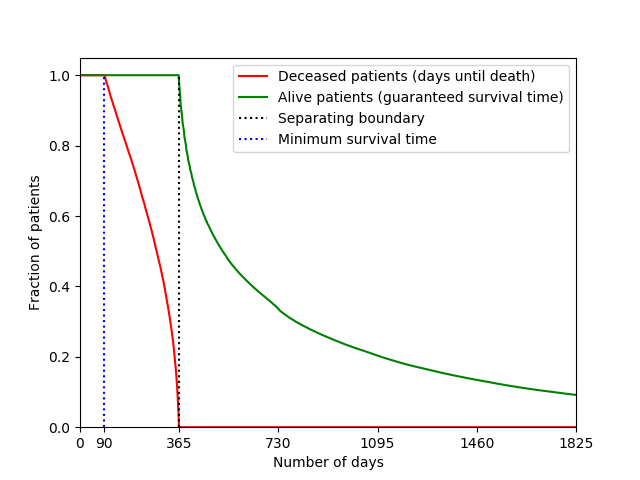}
    \caption{Right-censoring lengths shown as a survival plot. \zxie{Not sure whether Kaplan Meier plot would be good to include as well.}}
\label{fig:censor}
\end{figure}

\subsubsection*{Positive Cases}  
The constraints for positive cases were decided based on the rationale that palliative care is most beneficial if the referral occurs 3-12 months prior to death. Predicting mortality within 3 months is considered too late due to the preparatory time required to start palliative care in general. On the other hand, a lead time longer than 12 months is problematic because making accurate predictions over such a long time horizon is difficult, and more importantly, palliative care interventions are a limited resource that are best focused on more immediate needs. The \textit{prediction date} for positive cases must meet all the following constraints:
%
%

\begin{itemize}
\item The \textit{prediction date} must be a recorded date of encounter.
\item The \textit{prediction date} must be at least 3 months prior to date of death (otherwise death is too near).
\item The \textit{prediction date} can be at most 12 months prior to date of death (otherwise death is too far).
\item The \textit{prediction date} must be at least 12 months after the date of first encounter (otherwise the patient lacks sufficient history on which to base a prediction).
\item In-patient admissions are preferred over other admission types for the \textit{prediction date}, as long as they meet the previous constraints (since it is easier to start the palliative care conversation with them).
\item The \textit{prediction date} must be the earliest among the possible candidate dates subject to previous constraints.
\end{itemize}

\subsubsection*{Negative Cases} 
For negative cases (patients without a date of death), we require that the patient was alive for at least 12 months from the \textit{prediction date}.  We choose the \textit{prediction date} such that it satisfies all the following constraints:

%
%

\begin{itemize}
\item The \textit{prediction date} must be a recorded date of encounter.
\item The \textit{prediction date} must be at least 12 months prior to date of last encounter (to avoid ambiguity of death after date of EHR snapshot).
\item The \textit{prediction date} must be at least 12 months after the date of first encounter (otherwise insufficient history).
\item In-patient admissions are preferred over other encounter types for the  \textit{prediction date}, as long as they meet the previous constraints (to serve as controls for the admitted positive cases).
\item The \textit{prediction date} must be the latest among the possible candidate dates subject to previous constraints.
\end{itemize}

\subsubsection*{Admitted patients}
Those patients whose \textit{prediction date} corresponds to an in-patient admission are considered \textit{admitted patients}. Remaining patients are considered non-admitted (note that non-admitted patients could still have other recorded admissions in their history).  Further, for \textit{admitted patients}, their \textit{prediction date} it is re-adjusted by incrementing it to be the second day of admission. The rationale for doing this is that patient records are generally updated with the latest data (preliminary tests, diagnostics etc.) within 24 hours of admission, and the second day is better suited for making a more informed prediction. Note that the \textit{admitted patients} are a subset of the included patients (and NOT a separate set).

All the data after the \textit{prediction date} is censored in both the positive and negative cases. The KM-plot of censor lengths is shown in Fig \ref{fig:censor},  highlighting the separation between the two classes at 365 days.

\subsection*{Data Description}

\begin{table}
\begin{center}
\begin{tabular}{|l|r|r|r|}
\hline
       & Alive & Deceased  & Total\\ \hline
In EHR & 1,880,096 & 131,009 & 201,1105 \\ \hline
Selected & 205,571 & 15,713& 221,284 \\ \hline
Admitted & 9,648 & 1,131 & 10,779 \\ \hline
\end{tabular}
\newline
\newline
\caption{Breakdown of patient counts.}
\label{table:inclusion}
\end{center}
\end{table}

The inclusion criteria selected a total of 221,284 patients. Table ~\ref{table:inclusion} shows the breakdown of these patients based on inclusion and admission.

\begin{figure}
    \includegraphics[width=\columnwidth]{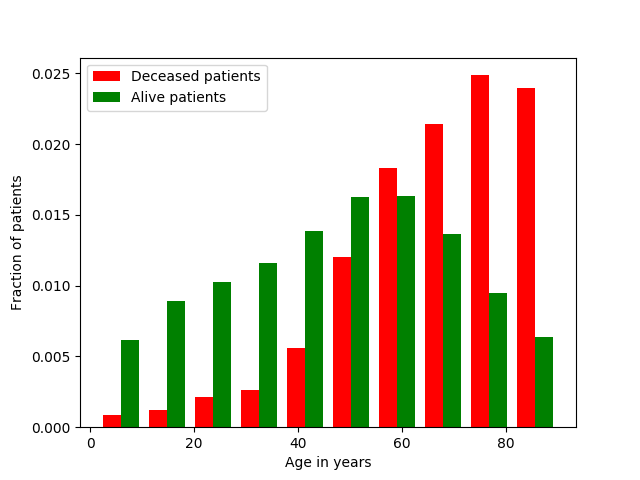}
    \caption{Age of patients at \textit{prediction time}. \fixed{If you have additional dimensions that could be plotted I would include some more.}}
    \label{fig:age}
\end{figure}

We observe that, unsurprisingly, the distribution of age at prediction time is not equal between the classes, and that the positive class (of deceased patients) is skewed towards older age (Fig ~\ref{fig:age}).

%
%

\begin{table}
\begin{center}
\begin{tabular}{|l|r|r|r|r|}
\hline
       & Training & Validation  & Testing & \\ \hline
Alive & 164,424 & 20,619 & 20,528 & 205,571\\ \hline
Deceased & 12,587 & 1,520 &  1,606 & 15,713\\ \hline
 & 177,011& 22,139 & 22,134 & 221,284\\
\hline
\end{tabular}
\newline
\newline
\caption{Data split for modeling.}
\label{tab:splits}
\end{center}
\end{table}

The included patients are randomly split in approximate ratio 8:1:1 into training, validation and test sets, as shown in Table \ref{tab:splits}. 

The prevalence of death among the included patients is approximately 7\%.  Approximately 5\% were \textit{admitted patients} (i.e., \textit{prediction date} was the second day of an admission). Among the \textit{admitted patients}, the prevalence of death is about 11\%. 

\subsection*{Feature Extraction}

For each patient, we consider the 12 months leading up to their \textit{prediction date} as their \textit{observation window}. Within the \textit{observation window} of each patient, we use ICD9 (International Classification of Diseases 9th rev) diagnostic and billing codes, CPT (Current Procedural Terminology) procedure codes, RxNorm prescription codes, and encounters found in that period to create features.

\begin{table}
\begin{center}
\begin{tabular}{|l|l|l|r|}
\hline
 & Start date & End date & Duration\\
\hline
\textit{observation window} & PD - 365 & PD & 365 \\
\hline
\textit{observation slice 1} & PD - 30 & PD & 30\\
\hline
\textit{observation slice 2} & PD - 90 & PD - 30 & 60 \\
\hline
\textit{observation slice 3} & PD - 180 & PD - 90 & 90 \\
\hline
\textit{observation slice 4} & PD - 365 & PD - 180 & 185 \\
\hline
\end{tabular}
\newline
\newline
\caption{Observation window and slices}
\label{tab:obswindow}
\end{center}
\end{table}

We create features as follows.  In order to capture the longitudinal nature of the data, we split the \textit{observation window} of each patient into four \textit{observation slices}, specified relative to the \textit{prediction date} (PD) as shown in Table \ref{tab:obswindow}

Thus, \textit{observation slice 1} is the most recent, and 4 is the oldest. The slice widths are intentionally uneven in order to give more emphasis to recent data. Within each \textit{observation slice}, we count the the number of occurrences of each code in each code category (prescription, billing, etc.) per patient. The count of every such code within the slice is considered a separate feature.

We also include the patient demographics (age, gender, race and ethnicity), and the following per-patient summary statistics in the \textit{observation window} for each code category:
\begin{itemize}
\item Count of unique codes in the category.
\item Count of total number of codes in the category.
\item Maximum number of codes assigned in any day.
\item Minimum number of codes (non-zero) assigned in any day.
\item Range of number of codes assigned in a day.
\item Mean of number of codes assigned in a day.
\item Variance in number of codes assigned in a day.
\end{itemize}

All these features (i.e, code counts in each of the four \textit{observation slices}, per category summary statistics over the \textit{observation window}, and demographics) were concatenated to form the candidate feature set. From this set, we pruned away those features which occur in 100 or fewer patients. This resulted in the final set of 13,654 features. Of the 13,654 features, each patient on average has 74 non-zero values (with a standard deviation of 62), and up to a maximum of 892 values. The overall feature matrix is approximately 99.5\% sparse.
%
%
%
%

\subsection*{Algorithm and Training}

Our model is a Deep Neural Network (DNN) \cite{LeCun2015DeepLearning} comprising an input layer (of 13,654 dimensions), 18 hidden layers (each 512 dimensions) and a scalar output layer. We employ the logistic loss function at the output layer and use the Scaled Exponential Linear Unit (SeLU) activation function \cite{Klambauer2017Self-NormalizingNetworks} at each layer. The model is optimized using the Adam optimizer \cite{Kingma2014Adam:Optimization}, with a mini-batch size of 128 examples. Intermediate model snapshots were taken every 250 mini-batch iterations, and the snapshot that performed best on the validation test was selected as the final model. Explicit regularization was not found necessary. The network configuration was reached by extensive hyperparameter search over various network depths (ranging from 2 to 32) and activation functions ($tanh$, $ReLU$ and $SeLU$).

\fixed{If you have GBM results may be nice to include as a baseline.}


The software was implemented using the Python programming language (version 2.7), PyTorch framework \cite{PyTorch}, and the scikit-learn library (version 0.17.1) \cite{scikit-learn}. The training was performed on an NVIDIA TitanX (12GB RAM) with CUDA version 8.0.

\subsection*{Evaluation}

Since the data is imbalanced (with 7\% prevalence), accuracy can be a poor evaluation metric \cite{HaiboHe2009LearningData}. The ROC curve can also be sometimes misleading on imbalanced problems\cite{DavisTheCurves} \cite{BoydUnachievableEvaluationb}. Therefore, we use the Average Precision (AP) score, also known as Area Under Precision-Recall Curve (AUPRC) for model selection \cite{Yuan2015Threshold-freeTests.}.


\section{ Results}

\begin{table*}[t]
\begin{center}

\begin{tabular}{|l|l|}
\hline
Patient MRN & XXXXXXX \\ \hline
Probability score & 0.946 \\ \hline
\end{tabular}
\newline
\begin{tabular}{|l|l|r|r|l|}
\hline
Factors & Code & Value & Influence & Description \\ \hline
Top Diagnostic factors              & V10.51    &         4   &   0.0051  &  Personal history of malignant neoplasm of bladder \\
              & V10.46     &        5    &  0.0019   & Personal history of malignant neoplasm of prostate \\
              & 518.5  &            1   &   0.0012   & Pulmonary insufficiency following trauma and surgery \\ 
              & 518.82   &          1    &  0.0008   & Other pulmonary insufficiency \\
               & 88.75    &         1    &  0.0006 &   Diagnostic ultrasound of urinary system \\ \hline
 Top Procedural factors             & 88331  &           1   &   0.0017  &  Pathology consultation during surgery with FS \\          
               &75984       &      1    &  0.0014   & Transcatheter Diagnostic Radiology Procedure \\            
               &72158       &      1   &   0.0013&    MRI and CT Scans of the Spine \\             
     &Code\_Type\_Count    &        76   &   0.0011&   Summary statistic (count of all ICD/CPT codes) \\    
              & 76005    &         1   &   0.0007&  Fluroscopic guidance and localization of needle or catheter tip for spine  \\      \hline
Top Medication factors& & & & \\ \hline
 Top Encounter factors           &Hx Scan    &        21   &   0.0012&  Number of scan encounters of all types \\
           &Inpatient     &       60     & 0.0004&  Number of days patient was admitted \\
   &Var\_Codes\_per\_Day    &         8    &  0.0002&  Summary statistic (variance in number of codes assigned per day)  \\
      &Code\_Day\_Count    &        88  &    0.0001&   Number of days any encounter code was assigned \\ \hline
   Top Demographic factors              & Age    &        81  &    0.0010& Age of patient in years at \textit{prediction time} \\ \hline
\end{tabular}
\newline
\newline
\caption{Prediction explanation generated on a random positive patient with high  probability score. Only factors that contributed to a drop in probability score are reported. \fixed{This table could use some formatting...tough to parse}}
\label{tab:explainpositive}
\end{center}
\end{table*}

\begin{table*}[t]
\centering
\begin{tabular}{|l|l|}
\hline
Patient MRN & YYYYYYY \\ \hline
Probability score & 0.909 \\ \hline
\end{tabular}
\newline
\begin{tabular}{|l|l|r|r|l|}
\hline
 Factors& Code & Value & Influence & Description \\ \hline
 Top Diagnostic factors              &197.7  &          16   &   0.1299 &  Malignant neoplasm of liver, secondary \\ 
               &154.1   &          3    &  0.1254&    Malignant neoplasm of rectum  \\ 
               &287.5    &         1  &    0.0194    &   Thrombocytopenia, unspecified \\        
               &780.6    &         1   &   0.0171&   Fever and other physiologic disturbances of temperature regulation \\
              &733.90    &         1   &   0.0113&   Other and unspecified disorders of bone and cartilage \\ \hline
 Top Procedural factors            & 73560  &           1    &  0.0502&    Diagnostic Radiology (Diagnostic Imaging) Procedures of the Lower Extremities \\
    & Code\_Type\_Count &           20   &   0.0491&     Summary statistic (Number of unique ICD-9/CPT codes) \\
          &     74160    &         1    &  0.0381&     Diagnostic Radiology (Diagnostic Imaging) Procedures of the Abdomen \\
  & Max\_Codes\_per\_Day &            6  &    0.0234&    Summary statistic (Maximum number of codes in any day) \\
& Range\_Codes\_per\_Day  &           6  &    0.0233&     Summary statistic (Range of codes across days) \\ \hline
  Top Medication factors       &     283838  &           1   &   0.0619&     Darbepoetin Alfa  \\
          &     28889   &          1    &  0.0247&     Loratadine \\ 
 &Range\_Codes\_per\_Day   &          5  &    0.0023&     Summary statistic (Ranges of codes across days) \\ 
   &Max\_Codes\_per\_Day   &          5  &    0.0023&     Summary statistic (Maximum number of codes in any day) \\
   &  Code\_Type\_Count    &         6   &   0.0015&     Summary statistic (Number of unique medication codes) \\ \hline
 Top Encounter factors            &Hx Scan       &     19   &   0.2239&     Number of scan encounters of all types \\
     & Code\_Day\_Count    &        97  &    0.0284&     Number of days any encounter code was assigned \\
       &   Outpatient      &      22    &  0.0228&     Number of Outpatient encounters \\
  & Var\_Codes\_per\_Day    &         1  &    0.0074&     Summary statistic (variance in number of codes assigned per day) \\ \hline
Top Demographic factors&     & & & \\ \hline
\end{tabular}
\newline
\newline
\caption{Prediction explanation generated on a random  false positive patient with high probability score. Only factors that contributed to a drop in probability score are reported.}
\label{tab:explainfp}
\end{table*}
\begin{figure}
    \includegraphics[width=\columnwidth]{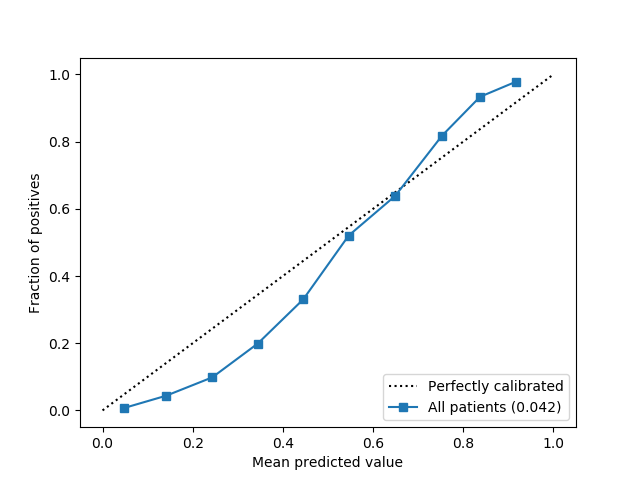}
    \caption{Reliability curve (calibration plot) of the model output probabilities on the test set data.}
\label{fig:calibration}
\end{figure}

\begin{figure}
    \includegraphics[width=\columnwidth]{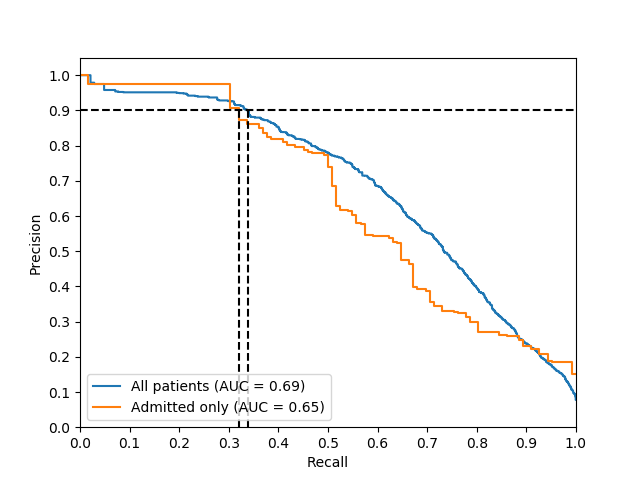}
    \caption{Interpolated Precision-Recall curve. The horizontal dotted line represents precision level of 0.9. The vertical dotted lines indicate the recall at which the curves achieve 0.9 precision.}
\label{fig:pr}
\end{figure}

\begin{figure}
    \includegraphics[width=\columnwidth]{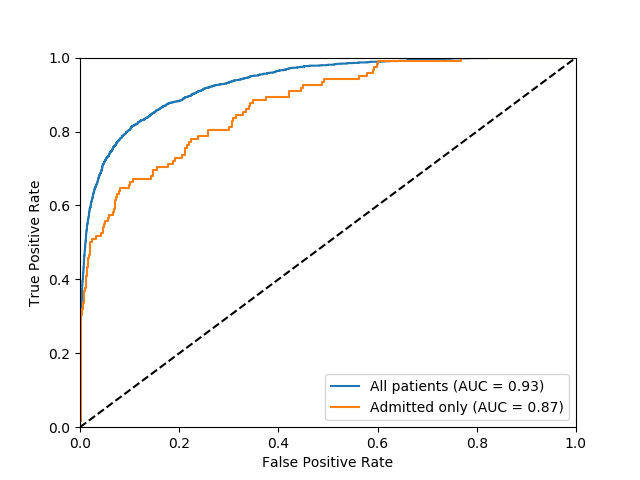}
    \caption{Receiver Operating Characteristic (ROC) of the model performance on the test set.}
\label{fig:roc}
\end{figure}

In this section we report technical evaluation results obtained on the test set using the model selected based on the best AP score on the validation set.

We observe that the model is reasonably calibrated (Fig \ref{fig:calibration}) with a \textbf{Brier score of 0.042}. In the high threshold regime, which is of interest to us, the model is a little conservative (under-confident) in its probability estimates, which should not hurt.

The interpolated Precision-Recall curve is shown in Fig \ref{fig:pr}. The model achieves an \textbf{AP score of 0.69} (0.65 on \textit{admitted patients}). Early recall is desirable, and therefore Recall at precision 0.9 is a metric of interest. The model achieves \textbf{recall of 0.34 at 0.9 precision} (0.32 on  \textit{admitted patients}). The Receiver Operating Characteristic curve is shown in Fig \ref{fig:roc}.  The model achieves an \textbf{AUROC of 0.93} (0.87 for \textit{admitted patients}). Both the ROC and Precision-Recall plots suggest that the model demonstrates strong early recall behavior. \fixed{Would be nice to have some qualitative statements about requirements of palliative care team and how the model exceeds these requirements and does well--baseline would also help for this purpose}

\subsection*{Qualitative Analysis}
\fixed{Would say more here or make a subsection / roll into another section}
It is worth recalling that predicting mortality was a proxy problem for identifying patients who could benefit from palliative care. In order to evaluate our performance on the original problem, we inspected false positives with high output probability.  Although such patients did not die within 12 months from their prediction dates, we noted that they were often diagnosed with terminal illness and/or are high utilizers of healthcare services. This can be seen in the positive and false positive examples shown in Section \ref{section:explain}. 

Upon conducting a chart review of 50 randomly chosen patients in the top 0.9 precision bracket of the test set, the palliative care team found all were appropriate for a referral on their \textit{prediction date}, even if they survived more than a year. This suggests that mortality prediction was a reasonable (and tractable) choice of a proxy problem to solve. 

\section{Explaining Predictions}
\label{section:explain}

Supervised machine learning techniques, and in particular Deep Learning techniques, have recently demonstrated tremendous success in predictive ability. However, better performance often requires larger, more complex models and thus sacrifice interpretability. It is worth drawing a distinction between interpreting a model, versus interpreting its decision \cite{RibeiroquotClassifierb} \cite{LiptonTheInterpretability}. While interpreting complex models (e.g very deep neural networks) may sometimes be infeasible, it is often the case that users only want an explanation for the prediction made by the model for a given example. It is important to establish the trust of the practitioner in the model's decisions for them to feel comfortable taking actions based on it. Providing explanations along with decisions help establish that trust.

We make the following observations to motivate our explanation technique.

\begin{itemize}
\item We can view the EHR data as a strictly growing log of events, and that new data is only added (nothing is modified or removed in general). This results in all our features being positive valued (as counts, means and variance of counts, etc).

\item We are most interested in explaining why a model assigns high probability to a patient. We are less interested in getting an explanation for why a healthy person was given a low probability (the reasons are also much less clear: the patient did not have brain cancer, did not have pneumonia, and so on). 

\item Directly perturbing feature vectors (e.g sensitivity analysis or for techniques described in \cite{RibeiroquotClassifierb}) does not work well in our case . For example, perturbing the feature representing the ICD count for brain cancer from zero to non zero can increase the probability of death significantly, implying that it is an important factor in general.  However, that is not a very useful observation for a \emph{specific patient who does not have brain cancer}. 

\end{itemize}

These observations motivate the following technique. For each ICD-9, CPT, RXNORM and Encounter code, we ablate \textit{all occurrences} of that code from the patient's EHR, create a new feature vector, and measure the drop in probability compared to the original probability.  This corresponds to asking: all else being equal, how would the probability change if this patient was not diagnosed with XYZ, prescribed drug ABC, etc? This drop in probability is considered the influence the code has on the model's decision for that patient. Demographic features are handled as follows. We zero out the age and  swap the gender to the opposite sex, and measure the respective drops in probability. Finally we sort the codes in descending order by influence, and pick the top 5 in each code category.  A random example of such a positive and false positive case are shown in Table \ref{tab:explainpositive} and \ref{tab:explainfp}.
\section{Conclusion}
We demonstrate that routinely collected EHR data can be used to create a system that prioritizes patients for follow up for palliative care \fixed{Seems stronger to say "We demonstrate that...". Also unsure if "model" or "system" would be better for this audience.}. In our preliminary analysis we find that it is possible to create a model for all-\fixed{typo}cause mortality prediction and use that outcome as a proxy for the need of a palliative care consultation\fixed{as a proxy for palliative care need?}. The resulting model is currently being piloted for daily, proactive outreach to newly admitted patients. We will collect objective outcome data (such as rates of palliative care consults, and rates of goals of care documentation) resulting from the use of our model \fixed{from the use of the model}. We also demonstrate a novel method of generating explanations from complex deep learning models that helps build confidence of practitioners to act on the recommendations of the system.

\section*{Acknowledgment}
We thank the Stanford Research IT team for their  support and help in this project. Research IT, and the Stanford Clinical Data Warehouse (CDW) are supported by the National Center for Research Resources and the National Center for Advancing Translational Sciences, National Institutes of Health, through grant UL1 TR001085. The content of studies done using the CDW is solely the responsibility of the authors and does not necessarily represent the official views of the NIH.

\bibliography{paper}
\bibliographystyle{IEEEtran}
%
%


\end{document}